# Controlled coupling of photonic crystal cavities using photochromic tuning

Tao Cai (蔡涛),[1,a)] Ranojoy Bose,[1,a)] Glenn S. Solomon,[2] and Edo Waks[1,2,b)]

[1]Department of Electrical Engineering, University of Maryland, College Park, MD 20742, U.S.A.
[2]Joint Quantum Institute, University of Maryland, College Park, MD 20742 and National Institute of Standards and Technology, Gaithersburg, MD 20899, U.S.A.

**Abstract**: We present a method to control the resonant coupling interaction in a coupled-cavity photonic crystal molecule by using a local and reversible photochromic tuning technique. We demonstrate the ability to tune both a two-cavity and a three-cavity photonic crystal molecule through the resonance condition by selectively tuning the individual cavities. Using this technique, we can quantitatively determine important parameters of the coupled-cavity system such as the photon tunneling rate. This method can be scaled to photonic crystal molecules with larger numbers of cavities, which provides a versatile method for studying strong interactions in coupled resonator arrays.

Photonic molecules consisting of two or more coupled microcavities are being explored for a variety of linear and nonlinear photonics applications such as biological and chemical sensors[1], optical memory[2], slow light engineering[3], and lasers[4]. Two-dimensional photonic crystals (PhC) offer an ideal device platform for realizing photonic molecules due to their inherently scalable planar architecture. PhC cavities also offer the capability of coupling to semiconductor quantum emitters such as quantum dots (QDs) in the strong coupling regime[5, 6], which enables strong

---

optical nonlinearities near the single photon level[7-11]. The interaction of photonic crystal molecules with quantum emitters could enable a broad range of applications such that include quantum computing[12], single photon generation[13,14], quantum-optical Josephson interferometry[15], and quantum simulation[16, 17].

Photonic crystal molecules have been experimentally realized in a number of previous works[4, 18-24]. In order to strongly couple PhC cavities, it is essential that their resonance frequencies be matched to within the normal-mode splitting. Once the detuning between the cavities is large compared to this splitting, they will decouple and behave as individual cavities as opposed to a coupled system. In PhCs, engineering coupled cavities with nearly identical frequencies is challenging because of fabrication inaccuracies. Previous studies of photonic crystal molecules overcame this problem by engineering large normal mode splitting exceeding 500 GHz[18-21]. For many applications, however, it is important to be able to accurately control the detuning between the cavities. This capability enables selective coupling and decoupling of cavity modes, which is important for controlling coupled-cavity interaction strength and also for characterizing fundamental physical properties such as photon tunneling rate. Accurate control of individual cavities in photonic crystal molecules could also enable novel reconfigurable photonic devices such as tunable filters[25, 26] and tunable lasers[27]. In addition, local tuning of cavities can serve to correct for fabrication imperfections, enabling the coupling of a large number of cavities to form complex arrays of coupled cavity structures[24].

A variety of methods have been demonstrated for tuning PhC cavity resonances. Nanofluidic tuning[22, 23] has been demonstrated as an effective room-temperature approach, but is difficult to apply for quantum optics applications that usually require low temperatures. Free-carrier injection[28] provides another approach for cavity tuning but typically provides only small resonance shifts and may also generate fluorescence from embedded quantum emitters. Thermo-optic tuning[29, 30] has been demonstrated in coupled cavity structures but is difficult to extend to arrays of closely packed devices and will also shift the resonance frequency of embedded quantum emitters such as QDs[31]. Other approaches utilize nano-mechanical tuning[32], which require a complex setup and can also strongly degrade the cavity quality factor ($Q$).

In this paper, we demonstrate local and reversible tuning of individual cavities in a photonic crystal molecule by using a photochromic thin-film[26,27]. The photochromic film is used to locally modify the effective index of refraction of individual cavities in the molecule, enabling them to be selectively coupled or decoupled. By tuning the cavities through the resonance condition, we observe clear normal mode splitting, enabling us to quantitatively determine the photon tunneling rate. We demonstrate the ability to resonantly tune both a two-cavity and a three-cavity photonic crystal molecule, which shows promise for scaling to more complex devices composed of large arrays of interacting cavities.

The device structure and calculated mode profile for a two-cavity photonic crystal molecule is shown in Figure 1(a) and (b). The device consists of two spatially separated linear defect L5 cavities[33,34]. In order to achieve high cavity $Q$s, the holes at the edges of the cavities were shifted by 0.196$a$ (labeled "A" in Figure 1(a)) and 0.046$a$ (labeled "B" in Figure 1(a)) respectively, where $a$ is the lattice constant of the triangular PhC structure. Here, $a$ is set to be 240 nm and the diameter of the holes is set to be 140 nm. The two cavities were separated by five rows of holes[18] as shown in Figure 1(a), corresponding to a center-to-center distance of 2.92 μm. The spatial mode profile of the device was obtained by finite-difference time-domain simulations and shows that the two-cavity photonic crystal molecule supports both symmetric and anti-symmetric modes with calculated $Q$ of $7\times10^5$ and $6\times10^5$ respectively, and a normal-mode splitting of 120 GHz (0.37 nm).

The designed device was fabricated using an initial wafer comprising of a 160-nm thick gallium arsenide (GaAs) membrane, grown on a 1-μm thick sacrificial layer of aluminum gallium arsenide ($Al_{0.78}Ga_{0.22}As$). A single layer of indium arsenide (InAs) QDs was grown at the center of the GaAs membrane (density 100-150 $QDs/\mu m^2$). The QDs served as an internal white-light source in order to optically characterize the device. A schematic of the device structure is shown in Figure 1(c). PhCs were defined on the GaAs membrane using electron-beam lithography and chlorine-based inductively coupled plasma dry etching. The sacrificial $Al_{0.78}Ga_{0.22}As$ layer was under-cut using a selective wet-etch process, leading to a free-standing GaAs membrane. A scanning electron microscope (SEM) image of a fabricated device is shown in Figure 1(d), with the two cavities labeled as C1 and C2 respectively. The photochromic thin-film layer was deposited on the structure through spin coating. Details of the photochromic film preparation and properties were previously reported[26, 27]. The film is composed of a mixture of 5 wt % 1,3,3-Trimethylindo linonaphthospirooxazine (TCI America) and 0.5 wt % 950 PMMA A4 dissolved in anisole. The solution was spun on the fabricated device surface at a spin rate of 3250 rpm, resulting in a film thickness of approximately 60 nm.

The fabricated device was mounted in a continuous flow liquid Helium cryostat and cooled to a temperature around 35K. QDs in the cavity regions were optically excited using a continuous wave Ti:Sapphire laser at 780 nm. Both cavities could be excited simultaneously by the excitation laser with a sufficiently large spot size or either singe cavities could be selectively excited by moving the laser spot around. The emission was collected using a confocal microscope with an objective lens (numerical aperture 0.7) and focused onto a pinhole aperture for spatial filtering. The aperture could be made large to collect emission from both cavities simultaneously, or it could be reduced to isolate the emission from only one of the two cavities. The collected emission was spectrally resolved using a grating spectrometer with wavelength resolution of 0.02 nm.

Figure 1(e) shows the photoluminescence (PL) spectrum (shown with green circles) of a fabricated device obtained by exciting both cavities simultaneously and collecting emission with a large pinhole aperture prior to tuning. The spectrum exhibits two bright peaks corresponding to the resonances of the two cavity modes (labeled CM1 and CM2 in the figure). The bright peaks were numerically fit to a double Lorentzian function, shown as blue solid line in Figure 1(e). From the fit, the resonant wavelengths of CM1 and CM2 were found to be 942.12 nm and 942.54 nm respectively, which corresponded to a spectral separation of 139 GHz (0.42 nm). The cavity linewidths were determined from the fit to be 0.09 nm and 0.04 nm respectively, corresponding to a cavity $Q$ of $1.05 \times 10^4$ ($\kappa/2\pi$ = 30 GHz) and $2.36 \times 10^4$ ($\kappa/2\pi$ = 14 GHz).

The sample was next illuminated with a focused ultraviolet (UV) laser emitting at 470 nm with an average intensity of 3 W/μm$^2$. The laser spot was focused on the cavity C1 and was sufficiently small to enable photochromic tuning of the CM1 resonance without affecting CM2.

Both cavities were excited with the 780-nm laser and the change in the cavity wavelength was monitored using PL emission collected from both of the cavity modes simultaneously using a large aperture. Figure 2(a) plots the PL emission intensity as a function of the detuning $\Delta = \lambda_{CM1} - \lambda_{CM2}$, where $\lambda_{CM1}$ and $\lambda_{CM2}$ represent the resonant wavelength of CM1 and CM2 respectively. Under UV exposure the wavelength of mode CM1 was red-shifted and became resonant with CM2 at a wavelength of 942.54 nm. As CM1 was tuned through CM2, a clear mode anti-crossing could be observed. A normal mode splitting of $\Omega_0/2\pi$ = 32GHz (0.10 nm) was measured when the two cavities were tuned on resonance, which can be used to calculate the photon tunneling rate $J$ between the two cavities using the equation[35]

$$J = \sqrt{\Omega_0^2 - [\omega_1 - \omega_2 - i(\kappa_1 - \kappa_2)]^2}/2 \qquad (1)$$

In the above equation, $\omega_1$ and $\omega_2$ are the angular frequencies of mode CM1 and CM2 respectively, while $\kappa_1$ and $\kappa_2$ are the individual cavity decay rates. From equation (1) it could be determined that $J/2\pi = 18$ GHz.

Figure 2(b) plots the linewidths of the two cavity modes as a function of detuning $\Delta$, where the linewidth was determined from the Lorentzian fit. As the cavities were brought into resonance, the linewidths of the two modes became identical, with both modes showing a fitted linewidth of 0.06 nm ($\kappa/2\pi$ =20 GHz; $Q = 1.57 \times 10^4$). Convergence of the linewidths is a strong evidence of hybridization of the two cavity modes into a pair of strongly coupled normal modes. As CM1 continued to be tuned through resonance, the modes reverted back to their original linewidths.

Figure 2(c) plots the intensities of the two modes as a function of detuning Δ. Mode CM1 initially exhibited a larger intensity prior to tuning, which was partly attributed to better overlap with the pump beam, resulting in stronger excitation of the QDs, as well as the fact that mode CM1 was centered on the aperture while CM2 was off-center resulting in a slightly lower collection efficiency. However, as CM1 was tuned on resonance with CM2, the cavity intensities became identical, exhibiting another signature of mode hybridization.

The observed results can be explained using a coupled-mode theory. We define the cavity field amplitudes as $a_1$ and $a_2$ corresponding to mode CM1 and CM2 respectively. The equations of motion for the two cavities can be described by the coupled-mode equations[36]

$$\frac{d}{dt}a_1(t) = -i\omega_1 a_1(t) - \frac{\kappa_1}{2}a_1(t) - iJa_2(t) - a_{int1}(t) \tag{2}$$

$$\frac{d}{dt}a_2(t) = -i\omega_2 a_2(t) - \frac{\kappa_2}{2}a_2(t) - iJa_1(t) - a_{int2}(t) \tag{3}$$

Here $a_{int1}(t)$ and $a_{int2}(t)$ are the cavity driving field amplitudes which may originate either from an external driving source, or in our case an internal source composed of excited QDs embedded in the membrane. Equation (2) and (3) can be solved in the frequency domain to arrive at

$$a_1(\omega) = \frac{-\kappa_2 a_{int1}(\omega)/2 + i(\Delta_2 a_{int1}(\omega) + Ja_{int2}(\omega))}{(i\Delta_2 - \kappa_2/2)(i\Delta_1 - \kappa_1/2) + J^2} \tag{4}$$

$$a_2(\omega) = \frac{-\kappa_1 a_{int2}(\omega)/2 + i(\Delta_1 a_{int2}(\omega) + Ja_{int1}(\omega))}{(i\Delta_2 - \kappa_2/2)(i\Delta_1 - \kappa_1/2) + J^2} \tag{5}$$

where $\omega$ is the driving field angular frequency, $\Delta_1 = \omega - \omega_1$, and $\Delta_2 = \omega - \omega_2$. The collected intensity from the two cavities (averaged over time) is given by,

$$\langle I(\omega) \rangle = \gamma_1 \langle |a_1(\omega)|^2 \rangle + \gamma_2 \langle |a_2(\omega)|^2 \rangle \tag{6}$$

where $\gamma_1$ and $\gamma_2$ are decay rates of the cavities into the collection mode (which is proportional to the collection efficiency). We assume that the driving sources at the two cavities are incoherent, which implies that $\langle a_{int1}(\omega) a_{int2}(\omega) \rangle = 0$. This assumption is highly realistic for our system because the cavities are driven by an inhomogeneous distribution of QDs that fluoresce independently, and each cavity is driven by different QDs.

The blue and green solid lines in Figure 2(b) and 2(c) plot the theoretically predicted behavior based on the coupled-mode theory. The inhomogeneous QD emission spectrum is much broader than both the cavity linewidths and the normal mode splitting. We therefore treat $a_{int1}(\omega) = a_{int1}(\omega_1)$ and $a_{int2}(\omega) = a_{int2}(\omega_2)$ as independent of $\omega$. We use $a_{int1}(\omega_1)$, $a_{int2}(\omega_2)$, $\gamma_1$ and $\gamma_2$ as fitting parameters. The calculated results using coupled-mode theory show extremely good agreement with the experimental results.

Figure 3 shows an example in which a two-cavity photonic crystal molecule was firstly red-shifted through resonance condition and then blue-shifted back to the resonant coupling point ($\Delta$ = 0nm), confirming the reversibility of the tuning method. This data was taken on a different device than the one used in Figure 2 but with an identical design. Figure 3(a) plots the PL spectra measured by selectively exciting CM1 (shown as blue circles) or CM2 (shown as green diamond) with the 780 nm laser and collecting light from both of the cavities simultaneously using a large aperture. Each spectrum exhibited only the resonance of the cavity being excited, indicating that CM1 and CM2 were initially decoupled due to large detuning. In the first step, CM1 was red-shifted through CM2 by focusing the UV laser and excitation laser on C1. Figure 3(b) shows the PL emission intensity as a function of detuning $\Delta$ as CM1 was tuned across the resonance with CM2. On resonance, a coupled mode with two peaks appeared in the spectrum by only exciting mode CM1, which indicated the resonant coupling point. Following the acquisition of the data in Figure 3(b), mode CM1 was red-shifted relative to CM2 due to photochromic tuning. A 532 nm green laser (intensity of 24 W/cm$^2$) was then utilized to blue-shift it back into resonance with CM2, as shown in Figure 3(c). Here, the ability to reversibly tune the cavity frequency plays an important role, enabling us to first find the resonant coupling point by red-shifting and then re-establishing it by reversing the shift.

Figure 4 shows results for photochromic tuning of a three-cavity photonic crystal molecule. Figure 4(a) shows the SEM image of a fabricated device, which consists of three linear defect L5 cavities (Labeled as C1, C2 and C3 respectively) spatially separated by five rows of holes. The three cavities have identical cavity designs to the ones described in Figure 1. All three cavity modes, labeled as CM1, CM2, and CM3, were initially detuned from each other. Mode CM2 was first tuned into resonance with CM1 to form the coupled modes CM$\pm$.

Figure 4(b) shows the PL spectra measured by selectively exciting C2 (shown as green diamonds) or C3 (shown as red circles) with the 780 nm laser and collecting light from all the three cavities simultaneously using a large aperture. When C2 was excited, the coupled mode spectrum CM$\pm$ was observed. However, when C3 was excited the spectrum showed only mode CM3 with little contribution from CM$\pm$, indicating that this mode was decoupled due to large detuning. By fitting the spectrum of CM$\pm$ to a double Lorentzian function (shown in Figure 4(b) as blue solid line), the two peaks of these coupled modes were determined to be located at 949.92 nm and 950.12 nm, with linewidths of 0.13 nm ($\kappa/2\pi$ = 43 GHz; $Q$ = 7.31×10$^3$) and 0.10 nm ($\kappa/2\pi$ = 33 GHz; $Q$ = 9.50×10³) respectively. The resonance of CM3 was fitted to a single Lorentzian function centered at 949.64 nm, with a linewidth of 0.10 nm ($\kappa/2\pi$ = 33 GHz; $Q$ = 9.50×10³).

The UV and excitation lasers were next focused only on C3 in order to red-shift mode CM3 into resonance with CM±. The PL emission intensity as a function of the detuning $\Delta' = \lambda_{CM3} - \lambda_{CM\pm}$ is plotted in Figure 4(c), where $\lambda_{CM3}$ represents the resonant wavelength of mode CM3 and $\lambda_{CM\pm} = (\lambda_{CM+} + \lambda_{CM-})/2$, where $\lambda_{CM+}$ and $\lambda_{CM-}$ are the resonant wavelengths of mode CM+ and CM- respectively. As CM3 was tuned near resonance with CM+ and CM- (red dashed box), a coupled mode with three peaks appeared in the spectrum by exciting only C3, which is a sign that all three modes are coupled.[37, 38] Figure 4(d) shows the three-peaked coupled modes spectrum. A triple Lorentzian fit was performed of the spectrum, where the three peaks of the coupled modes were determined to be centered at 949.82 nm, 949.98 nm and 950.16 nm with linewidth of 0.11 nm ($\kappa/2\pi$ = 37 GHz; $Q$ = 8.63×10³), 0.11 nm ($\kappa/2\pi$ = 37 GHz; $Q$ = 8.63×10³) and 0.12 nm ($\kappa/2\pi$ = 40 GHz; $Q$ = 7.92×10³).

In conclusion, we have presented a technique to control the coupling interaction between individual cavities in a photonic crystal molecule by tuning with a photochromic thin-film. This approach could be applied to photonic crystal molecules with larger numbers of cavities, as well as hetero-structures such as cavity-waveguide systems. The method is highly versatile and could also be applied to other cavity architectures such as micro-disk resonators[39] and micro-ring resonators[40]. Ultimately, these results could pave the way for development of complex and highly reconfigurable integrated photonic devices composed of a large array of nanophotonic cavities.

**Acknowledgements**

T. Cai and R. Bose contributed equally to this work. The authors would like to acknowledge support from a DARPA Defense Science Office Grant (No. W31P4Q0910013), the ARO MURI on Hybrid quantum interactions Grant (No. W911NF09104), the physics frontier center at the joint quantum institute, and the ONR applied electromagnetic center N00014-09-1-1190. E. Waks would like to acknowledge support from an NSF CAREER award Grant (No. ECCS-0846494).

**Figures**

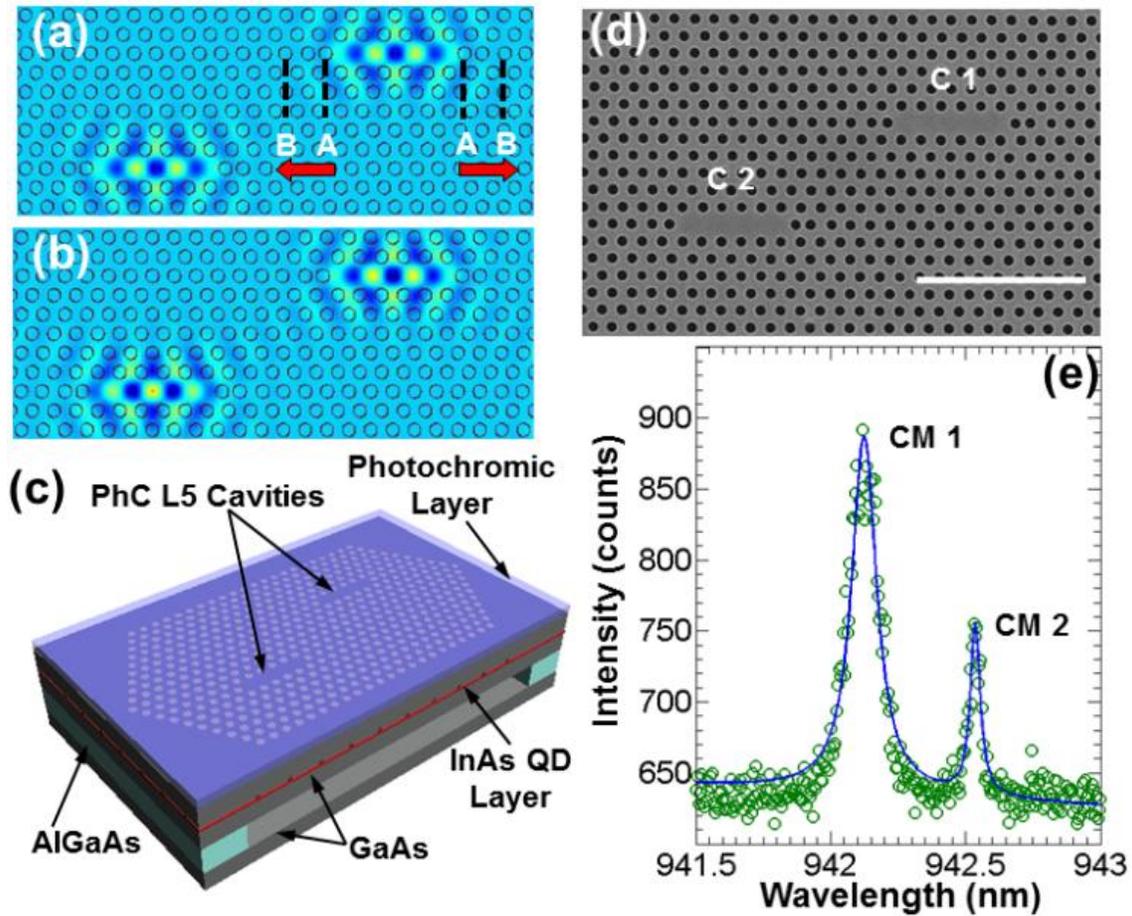

FIG. 1. (Color) Calculated mode field profile ($E_y$) for (a) symmetric and (b) anti-symmetric modes. A and B label holes shifted to improve cavity $Q$. (c) Three-dimensional schematic layout of the designed device. (d) SEM image of fabricated two-cavity photonic crystal molecule. Scale bar: 2 $\mu m$. (e) Measured PL spectrum of two cavity modes corresponding to individual cavity resonances shown with green circles. Double Lorentzian fit shown as blue solid line.

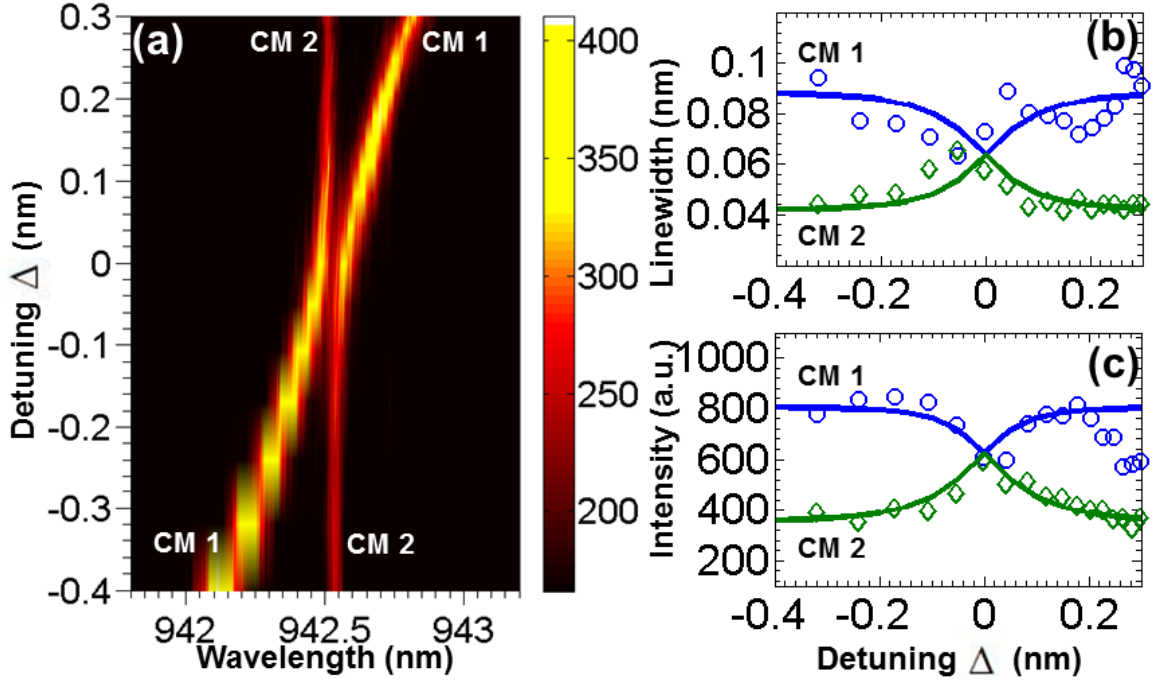

FIG. 2. (Color) (a) PL emission intensity as a function of the detuning $\Delta$ between the two cavities. (b) Measured linewidths of CM1 (shown as blue circles) and CM2 (shown as green diamonds) as a function of detuning $\Delta$. Theoretical fits based on coupled-mode theory are shown as blue and green solid lines. (c) Measured intensities of CM1 (shown as blue circles) and CM2 (shown as green diamonds) as a function of detuning $\Delta$, along with theoretical fits shown as blue and green solid lines.

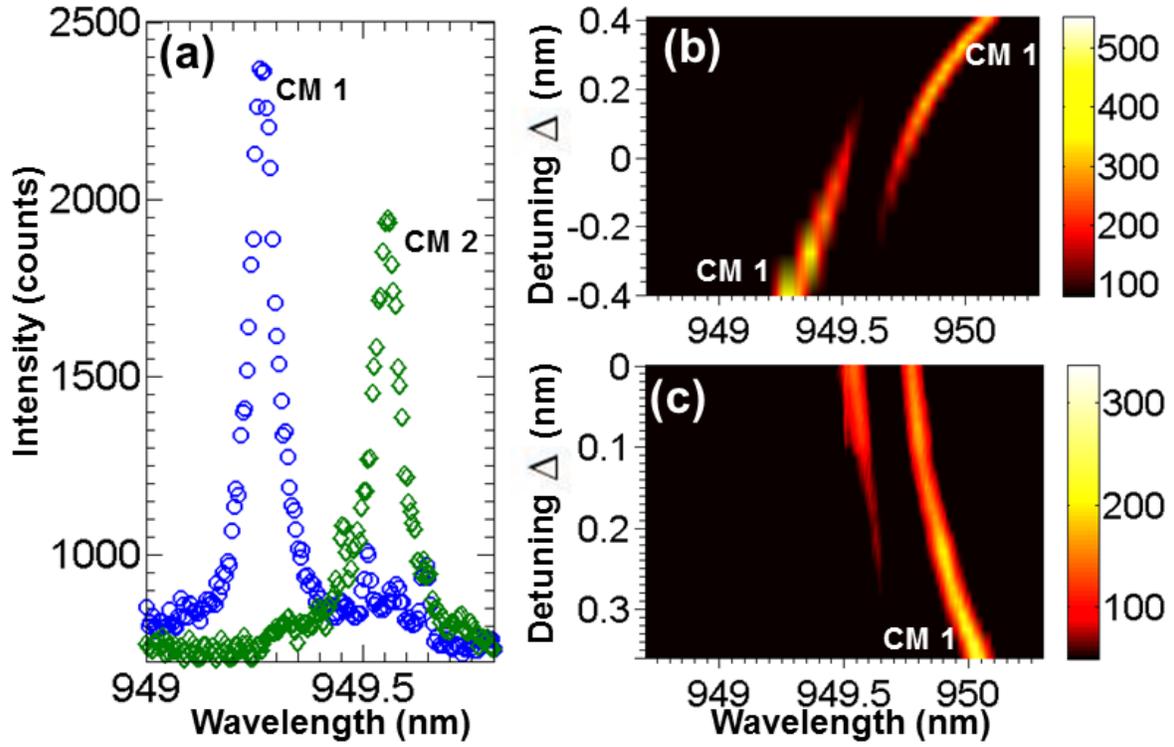

FIG. 3. (Color) (a) Measured spectra of decoupled CM1 (shown as blue circles) and CM2 (shown as green diamonds). The PL emission intensity as a function of the detuning $\Delta$ is shown for both (b) under UV illumination and (c) visible (532 nm) light illumination.

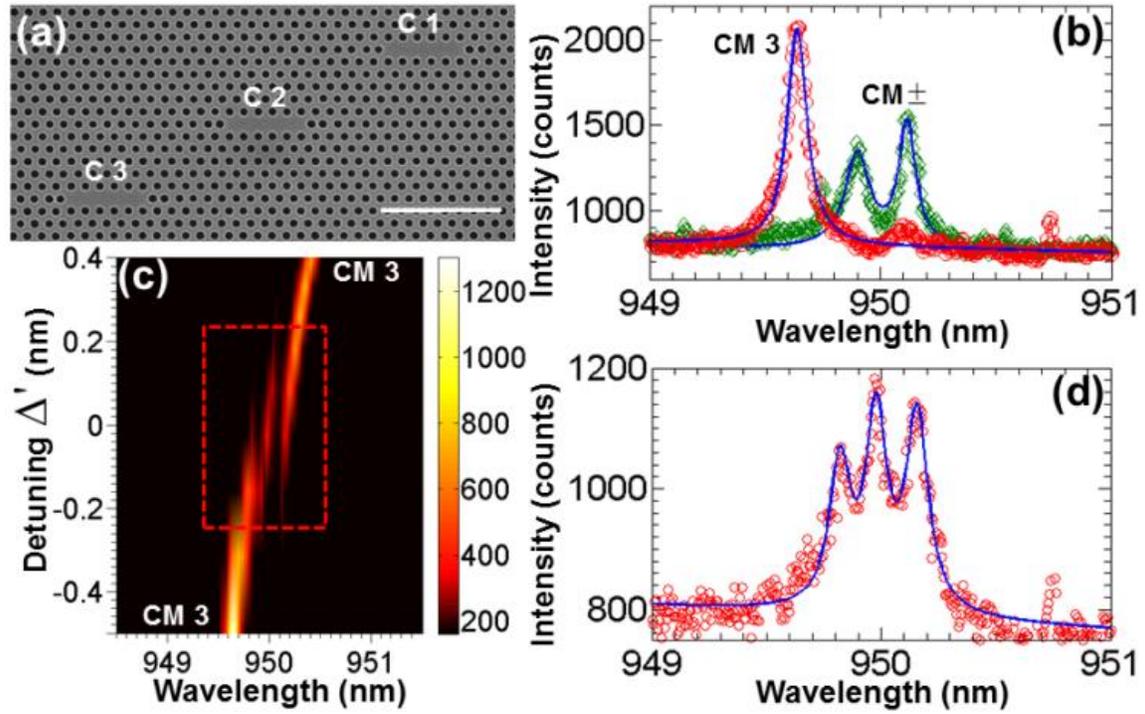

FIG. 4. (Color) (a) SEM image of fabricated three-cavity photonic crystal molecule. Scale bar: 2 $\mu m$. (b) Measured PL spectra of coupled modes CM± shown with green diamonds and decoupled CM3 shown with red circles. Solid lines show fit to Lorentzian functions. (c) PL emission intensity as a function of detuning $\Delta'$. The rectangular area denoted by the red dashed line indicates the resonant interaction regime where a three-peaked triplet was observed. (d) Measured PL spectrum taken at $\Delta'=0$. Lorentzian fit shown as blue solid line.